\documentclass[twocolumn]{revtex4}
\usepackage[dvips]{graphicx}
\begin{document}

\title{Design of magnetic tweezers for DNA manipulation}

\author{Doriano Brogioli}
\email{dbrogioli@gmail.com}

\affiliation{Dipartimento di Medicina Sperimentale, Universit\`a degli Studi di
Milano - Bicocca, Via Cadore 48, Monza (MI) 20052, Italy.}

\begin{abstract}
We study different configurations of permanent magnets and ferromagnetic 
circuit, in order to optimize the magnetic field for the so-called
``magnetic tweezers'' technique, for studing mechanical properties of DNA molecules.
The magnetic field is used to pull and twist a micron-sized superparamagnetic bead,
tethered to a microscope slide surface by a DNA molecule.
The force applied to the bead must be vertical, pointing upwards, being as strong as possible, and 
it must decrease smoothly as the magnets are moved away from the bead. In order to rotate 
the bead around the vertical axis, the field must be horizontal. Moreover, the volume
occupied by the magnets is limited by the optical system.
We simulate different configurations by solving the equations for the static magnetic field;
then, we test some of the configurations by measuring the force acting on a bead tethered
by a DNA molecule. One of the configurations is able to generate a magnetic field ten times
stronger than usually reported.
\end{abstract}

\maketitle

\emph{A closely related paper \cite{lipfert2009} has been recently published on the same topic!}

The so-called ``magnetic tweezers'' technique allows to manipulate 
small bodies or macromolecules, by assembling the sample with 
a micron-sized superparamagnetic 
bead. Manipulation is performed by applying a force to the bead, through a 
magnetic field. \cite{crick1950,smith1992,wirtz1995,uchida1998,
strick1998,leger1999,guthold1999,wang1999}

In particular, DNA manipulation is performed by tethering a superparamagnetic bead
to a microscope slide through a DNA molecule. The position of the bead, as a function
of the magnetic force, gives interesting informations on the mechanical properties
of the molecules \cite{smith1992, strick1998} (see Ref. \cite{strick2003} for 
a review, and references therein).

In this note, we focus on the structure of the magnets and magnetic circuit
of an instrument with permanent magnets, like the one described in Ref. \cite{strick1998bis},
with the purpose of optimizing the magnetic field. The field
we want to obtain has the following properties.
\begin{enumerate}
\item The force it generates on the superparamagnetic bead is vertical and pointing upwards.
\item The force intensity can be accurately changed by moving the magnets.
\item The maximum value is as strong as possible.
\item The field prevents the bead to rotate freely around the vertical axis, and 
can impose a rotation of a given angle.
\end{enumerate}

Moreover,  the volume occupied by the magnets is limited by the optical system.
The DNA sample and the bead are on a microscope slide. 
The volume under the slide is occupied by the microscope objective; the magnets
must occupy the volume above the microscope slide.

Some of magnetic tweezers use magnetic coils. \cite{haber2000}
This gives some advantages, the main one being
the possibility to change the force by changing the current in the coils. Unfortunately, 
this setup is quite big, so that it is difficult to move the magnetic circuit. On the contrary,
permanent magnets allow the contruction of compact magnetic circuits, that can be rotated 
around the vertical axis; this is useful when we want to rotate the bead, in order to twist 
the DNA molecule.

\begin{figure}
\includegraphics[angle=-90,scale=0.97]{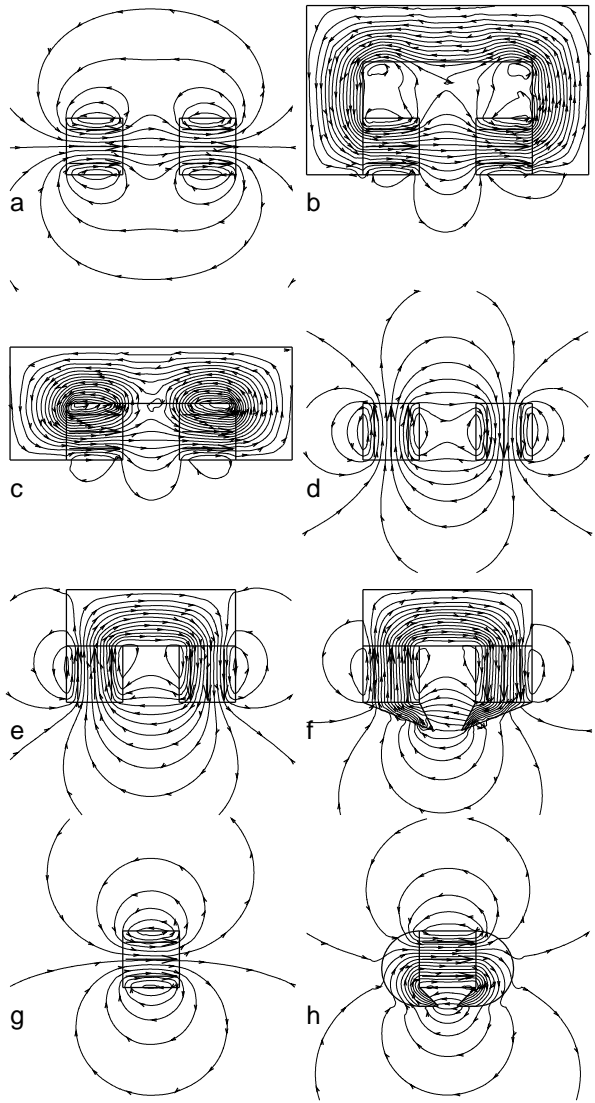}
\caption{Magnet configurations. The thick lines are the borders of 
ferromagnetic parts; the areas with thick arrows are the permanent magnets,
with arrows representing permanent magnetzation.
The sample is below the magnetic circuit, on the axis. Thin lines represent
the field lines we evaluated numerically.}
\label{fig_schemi}
\end{figure}

In order to find a good configuration of the permanent magnets, magnetic circuit 
and polar expansions,
we analyzed different configuration of magnets and ferromagnetic polar expansions,
shown in Figs. \ref{fig_schemi}.

For each configuration, we calculated numerically the magnetic field, by 
solving the equations for static magnetic field $\mathrm{div}\vec{B}=0$ and 
$\vec{\mathrm{rot}}\vec{H}=0$. The phenomenological relation between $\vec{B}$ and 
$\vec{H}$ is linear in air: $\vec{B} = \mu_0 \vec{H}$. In the ferromagnetic material,
we assume a non-linear relation, with a saturation at $B_{sat}$, that interpolates
between the two limit cases:
\begin{equation}
\begin{array}{lr} 
\vec{B} = \mu_R \mu_0 \vec{H} 
&
\left|\vec{H}\right|\ll \frac{B_{sat}}{\mu_r\mu_0} \\
\vec{B} = B_{sat} \frac{\vec{H}}{\left|\vec{H}\right|}  
+ \mu_0 \vec{H}
&
\left|\vec{H}\right|\gg \frac{B_{sat}}{\mu_r\mu_0} \\
\end{array}
\end{equation}
In permanent magnets, we assume 
$\vec{B} = \mu_0 \left( \vec{H} + \vec{H_0} \right)$.

The equations are solved numerically, with a finite element method, 
by using the program FreeFEM \cite{freefem}.

We assume translational simmetry, so that we work in a 2d system.
All simulation are done whit the following parameters: 
magnetization of the magnets  $H_0 = 10^6  \mathrm{A m^{-1}}$; 
saturation field for the ferromagnet  $B_{sat} = 1.7 \mathrm{T}$;
permeability of the ferromagnet for low fields $\mu_R = 5000$.
The side of the square permanent magnets $L$ will be taken as length unit.

The field lines we calculated are reported in Fig. \ref{fig_schemi}.

The magnetic force acting on a paramagnetic object is
$ \vec{F}= \mu \vec{\nabla}\left(\vec{B}^2\right) V$, 
where $\vec{B}$ is the magnetic induction field, $\mu$ the diamagnetic constant
and $V$ the volume of the bead.
To maximize $\vec{F}$, once we have fixed the type of material of the bead and so $\mu$, 
we have to maximize $\vec{\nabla}\left(\vec{B}^2\right)$. 
In order to obtain a strong force directed upwards, the magnetic field must
have a fast increase in the same direction. 
Moreover, the direction of the field must be horizontal, so that 
the rotation of the magnets around vertical axis results in a rotation of the 
bead.

\begin{figure}
\includegraphics[angle=-90]{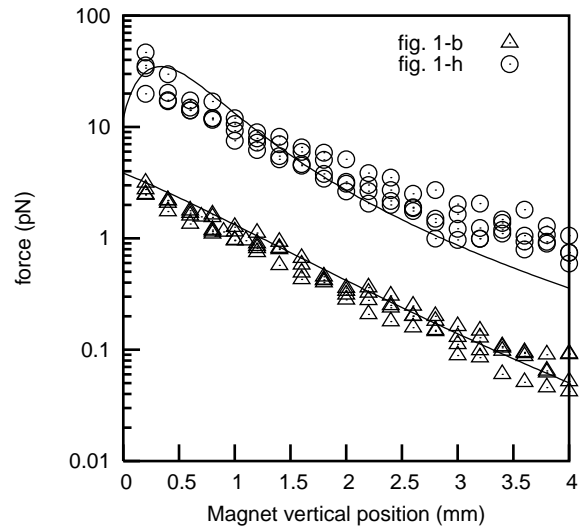}
\caption{Comparison between the calculated (lines) and experimentally measured forces (points).
}
\label{fig_forces}
\end{figure}

In each of the configurations shown in Fig. \ref{fig_schemi}, due to the symmetry
of the problem, the field is horizontal along the symmetry axis, where the sample is placed, as
required. In the figures, the density of field lines is proportional to the magnetic field;
this means that the force acting on the superparamagnetic bead is directed towards
regions with higher density of field lines; in the configurations we
studied, it is always vertical on the axis, due to the simmetry of the problem.
We calculated the force as a function of $z$, the distance
between the sample and the lower part of the magnets or of the polar expansions.
In Fig. \ref{fig_forces} we report the forces we calculated for the configurations B and H;
the force refers to the $2.8\mathrm{\mu m}$ diameter beads we used in the experiments.


The configurations in Figs. \ref{fig_schemi}-a, b and c have two magnets, with horizontal, 
parallel magnetizations; they are aligned, with a small gap between north and south faces.

The configuration in Fig. \ref{fig_schemi}-a
shows the two magnets standing alone. All the magnetic flux that comes from the left magnet
entres in the right one, but the presence of the gap allows the magnetic field to spread 
into the space below the magnets, where the sample will be placed. Looking at the 
force along the axis of symmetry, we can divide the space below the magnets in consecutive 
regions. In the first one, going downwards from the edge of the magntes,
there's a strong force directed upwards (from $2.4\mathrm{pN}$ near the edge to $0$)
due to the strong decrease of the magnetic field, mainly conveyed into the gap;
the density of these lines decreases going away from the surface 
of magnets.
Further away from the magnets, the magnetic field changes its direction, due to the flux that goes 
in the opposite direction and connects the faces more far apart of the two magnets.
At the point of inversion
the force vanishes; below this point, there's a region where the force is very slightly repulsive
(from $0$ to $0.04\mathrm{pN}$), 
that is, directed downwards, and then the force becomes again attractive, but very weak
(from $0$ to $0.02\mathrm{pN}$, and again to $0$ when
we go to infinity).

The main defect of this configuration is that the force vanishes abruptly when going 
far from the magnets; the sample must be very close to the magnets and, in order
to control the force, very small errors in the position of the magnets generate 
strong errors in the force. Moreover, near the region where the 
vertical component of the force vanishes, horizontal components take a stronger
relative influence.

In Fig. \ref{fig_schemi}-b, we added a ferroelectric ring that closes the magnetic
circuit between the faces of the magnets most far apart. This enhances the magnetic field,
but also avoids the inversion of the field. In this 
configuration the force ranges from $5\mathrm{pN}$ near the magnets to $0$ as we go to 
infinity. 

Figure \ref{fig_schemi}-c is a simplified version of the previous configuration, where
the ferroelectric ring passes near the gap between the magnets. We can see that some of the
flux is conveyed into the ferroelectric, and is thus lost. The force start from the 
maximum value $3.6\mathrm{pN}$ and goes to $0$ as we go to infinity. This shows the importance of
avoiding ferroelectric parts near the gap of the magnets.


The configurations in Figs. \ref{fig_schemi}-d, \ref{fig_schemi}-e and \ref{fig_schemi}-f
have two magnets, with anti parallel vertical magnetizations, 
aligned horizontally, with a small gap between them.

In Fig. \ref{fig_schemi}-d the two magnets are alone, without ferromagnetic circuit.
The field lines that come from one side of the magnet can close on the other magnet, or on the
opposite side of the same magnet. Along the axis of simmentry there's only one point where the
magnetic field lines have a maximum of density, and we 
distinguish two regions along the symmetry axis. 
In the first one, from the edge of the magnets until a distance of $0.12 L$, the force is
slightly repulsive (from $0.9\mathrm{pN}$ near the magnets to $0$).
In the second, from $0.12 L$ to infinity, the force has a maximum value
$0.9\mathrm{pN}$ at a distace $0.4 L$ from the magnets, and then goes slowly to zero.

The configuration in Fig. \ref{fig_schemi}-e is improved with respect to 
the previous one, because we prevent 
the field lines, generated from a magnet, to close to the other side of the same magnet.
The result is similar to case D, but with 
stronger force ($1.8\mathrm{pN}$ at a ditance $0.84 L$ from the magnets).

Figure \ref{fig_schemi}-f shows the effect of polar expansions, where no inversion of force is present;
the force starts from $6\mathrm{pN}$ at the end of the tips, then has a maximum
value $9\mathrm{pN}$ at a distance $0.1 L$ from the tips' end; then it goes to zero for increasing distance.


The configurations in Figs. \ref{fig_schemi}-g and \ref{fig_schemi}-h 
have only one magnet with horizontal polarization.

Figure \ref{fig_schemi}-g shows the field generated by a single magnet.
Since the force is generated by the field lines that close between the two opposite,
quite far, sides of the magnet, both the field and the force are quite small.

In Fig. \ref{fig_schemi}-h we show the effect of polar expansions,
that convey the great part of the flux.
We decide to drive the magnetic field lines using
ferromagnetic polar expansion that can drive most of the field above the sample;
we will use tips at the end of the polar expansion to generate a strong gradient of $\vec{B}^2$.
As the field lines are conveyed into smaller section of the polar expansion,
their density is increases, until the field reaches the saturation value.
After this value, the field lines spread out of the polar expansion creating a strong gradient.
In this case we have very close tips with a very short path in air for the field lines, resulting in 
a very strong force ($50\mathrm{pN}$ at a distence of $0.05 L$ from the tips' end)


The experimental setup and the data analysis closely mirrors the
procedures described in Ref. \cite{strick1998bis, strick1998}; in particular, we are 
interested in measuring the forces generated by the magnet, by observing the amplitude of
movement of the bead.

The magnets we used are $6\mathrm{mm}\times 6\mathrm{mm}\times 5\mathrm{mm}$
long, with magnetization parallel to the shorter side. The polar expansions and
the rings composing the magnetic circuit are made of soft iron.

In Fig. \ref{fig_forces} we report the forces we measured; the experimental
points represent six different consecutive scan of the position of the magnets for the configuration
in Fig. \ref{fig_schemi}-b, and four for the configuration in Fig. \ref{fig_schemi}-h.
In both the configurations, the permanent magnets and the microsphere were the same
($2.8\mathrm{\mu m}$ Dynabeads, streptavidin coated).

We tried the three configuration of Figs. \ref{fig_schemi}-a, \ref{fig_schemi}-b and 
\ref{fig_schemi}-c; we obtained good results only for the case \ref{fig_schemi}-b.
The force we obtained was roughly comparable to the ones 
reported in literature, though an accurate comparison should require the knowledge
of the magnetization of the permanent magnets and of the volume and magnetic permeability 
of the microsphere.

The configuration in Fig. \ref{fig_schemi}-h proved to be the most 
efficient. Forces up to 10 times higher than the ones previously reported 
for permanent magnet setups \cite{strick1998bis} have been obtained.

We thank Roberto Ziano for useful discussions.

\bibliography{magneti.bib}

\end{document}